# Could ART increase the population level incidence of TB?

Brian G. Williams

South African Centre for Epidemiological Modelling and Analysis (SACEMA), Stellenbosch, South Africa

Correspondence to BrianGerardWilliams@gmail.com

**Abstract**

HIV increases the likelihood that a person will develop TB. Starting them on anti-retroviral therapy (ART) reduces their risk of TB but not to the level in HIV negative people. Since HIV-positive people who are on ART can expect to live a normal life for several decades this raises the possibility that their elevated risk of infection, lasting for a long time, could lead to an increase in the population level incidence of TB. Here we investigate the conditions under which this could happen and show that provided HIV-positive people start ART when their CD4$^+$ cell count is greater than 350/μL and that there is high coverage, ART will not lead to a long-term increase in HIV. Only if people start ART very late and there is low coverage of ART might starting people on ART increase the population level incidence of TB.

## Introduction

Detailed epidemiological models have shown that TasP, or 'treatment-as-prevention', can eliminate HIV[1] and HIV-related TB.[2] When people are infected with HIV their risk of developing TB increases exponentially with declining CD4$^+$ T-lymphocyte cell counts (hereafter 'CD4 count').[3] When such people are on anti-retroviral therapy (ART) their CD4 counts recover over a period of four to five years[4-6] and the risk of developing TB falls by about 60% at all CD4 counts.[2] Nevertheless, because the incidence of TB in HIV-positive people on ART remains higher than it is in HIV-negative people and people on ART should live for a long time, the possibility exists that under some circumstances ART could have the paradoxical effect of increasing the overall incidence of TB. Here we explore the conditions under which this might happen.

## Methods and assumptions

We assume:

1. That the median CD4 count in HIV negative people is 1,200/μL; observed values range widely[6,7] from 1,150/μL[8] in a study in Uganda to 591/μL in a study in Botswana;[9]
2. That the CD4 count drops by 25% immediately after the acute phase of infection,[7] that is about one month after infection with HIV;[10]
3. That after the acute phase of infection the CD4 counts falls linearly to zero,[7] when the person dies if they do not receive ART;
4. That the median life expectancy for HIV positive people, without ART, is ten years[6,11] and with ART is forty years;[12]
5. The incidence of TB in HIV-positive people increases by 38% (25%−53%) for each decrease of 100 cells per μL[3] as the HIV-infection progresses, based on studies in Italy[13] and in Cape Town (Figure 1);[14]
6. When people start ART their CD4 count converges exponentially with time to an asymptote that is 371/μL (362/μL−381/μL) above the value at which they start ART at a rate of 0.60 (0.55−0.66) per year.[4,6] This increase in CD4 counts implies a reduction in TB incidence of 76% (60%−85%).* Direct estimates of TB comparing people on and off ART suggest a reduction of 61% (54%−68%).[2]

We consider a person infected with HIV, calculate the relative risk of developing TB as their CD4 counts decline with time since infection and then as their CD4 counts increase once they are on ART. We assume that people who start ART, are fully adherent with treatment, and have good viral load suppression will live for 40 years.[12]

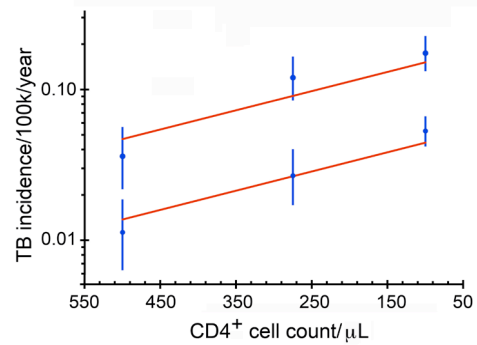

Figure 1. TB incidence as a function of CD4 counts in HIV-positive people from studies in public hospitals in Italy[13] and South Africa.[14] The fitted lines have a slope of 0.0038 (0.0025−0.0053) μL.

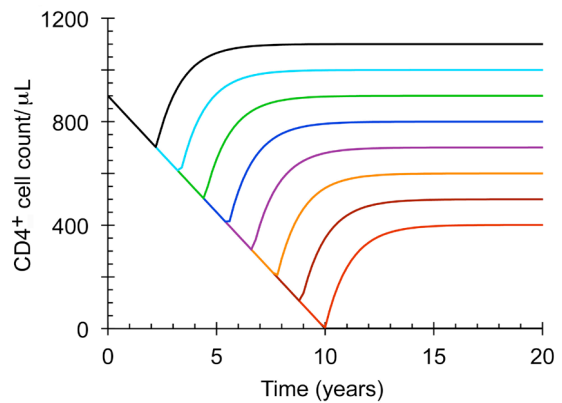

Figure 2. The decline and recovery of the CD4$^+$ T-cell count after infection with ART assuming that people start treatment at different CD4 counts. The lines, from the bottom (red) to the top (brown), correspond to starting CD4 counts of 0, 100, 200, … 700/μL.

---

\* Risk reduction = $1- e^{-371 \times 0.0038}$

## Results

The decline and the recovery of CD4 counts is shown in Figure 2 and the incidence of TB is shown in Figure 3, in both cases as a function of the time since infection with HIV and the CD4 count at which people start ART. From the data in Figure 3 we calculate the lifetime risk of TB as a function of the CD4 counts at the start of treatment (Figure 4); the red dots and lines are for a person who never starts ART and dies after ten years. The intersection of the red line with the horizontal axis gives the value of the CD4 count at which the lifetime risk of developing TB is the same with and without ART. This CD4 count defines a threshold above which ART reduces the lifetime risk of TB and below which it increases the lifetime risk of TB.

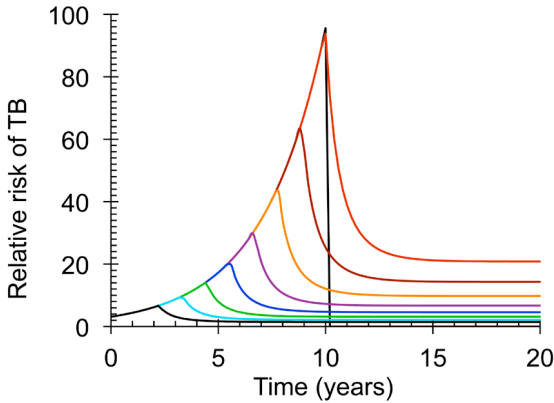

Figure 3. The relative risk of TB as a function of the CD4 count at the start of treatment and the time since infection. The reference value of 1 corresponds to the incidence of TB in HIV-negative people for the same population. The lines from the top (red) to the bottom (brown) correspond to starting ART at CD4 counts of 0, 100, 200, … 700/µL. The incidence of TB increases at a rate of 38% for each drop of 100 cells/µL in the CD4 count.

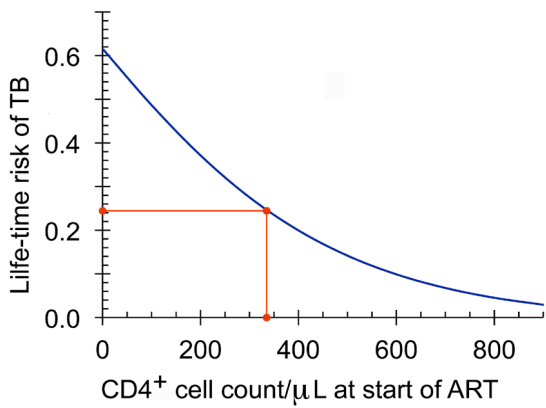

Figure 4. The lifetime risk of TB as a function of the CD4 count at the start of treatment. The red dots and line are for a person who never starts treatment and dies ten years after they are infected.

If a person has a CD4 count of 1,200/µL, never starts ART and dies after ten years, and the risk of TB increases at 38%/µL, their lifetime risk of developing TB is 24% (black line, Figure 3; red dots, Figure 3Figure 4). Provided they start ART at a CD4 count greater than 335/µL their lifetime risk of TB is less than if they never started treatment and vice versa (Figure 4).

The two key parameters in the model are the CD4 count, before the person becomes infected with HIV, and the rate of increase in the incidence of TB as the CD4 count declines. To investigate the sensitivity of the CD4 threshold to these parameters we vary the mean CD4 count in HIV-negative people from 600/µL to 1,200/µL. Reducing the initial CD4 count reduces the both the lifetime risk at any CD4 count, the life-time risk without treatment, and the critical value of the CD4 count, below which treatment increases the life-time risk of TB. Because they all vary in a similar way the CD4 threshold only varies from 338/µL to 371/µL as shown in Figure 5.

Survival without ART is sensitive to age and can be as long as 16 years for those infected at age 4 years to 4 years for those infected at age 65 years.[15] However, the CD4 threshold is also insensitive to changing the life expectancy from ten years to five or fifteen years. Doubling the life expectancy, say, reduces the rate at which CD4 counts decline by one-half, which is equivalent to halving the initial CD4 count, and this does not change the threshold CD4 significantly.

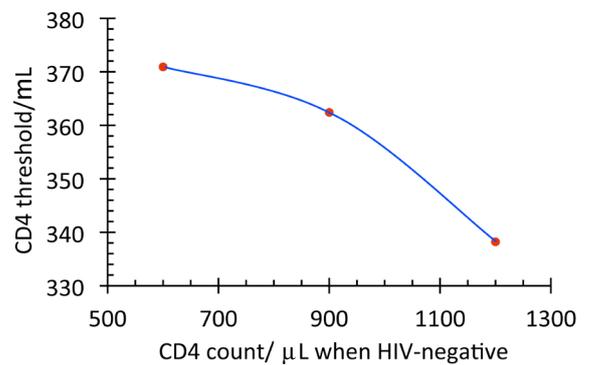

Figure 5. The CD4 threshold, below which ART will lead to an increase in the incidence of TB, as a function of the CD4 counts in HIV-negative people.

However, changing the rate at which the incidence of TB increases as the CD4 count falls has a much greater effect on the CD4 threshold. If we set the CD4 count in HIV-negative people to 900/µL, and set the rate parameter to the lower limit of 0.0025 µL/cell, the lifetime risk of TB without treatment falls to 8% and CD4 threshold increases to 560/µL. Setting the rate parameter to the upper limit of 0.0053 µL/cell the lifetime risk of TB without treatment increases to 72% and CD4 threshold falls to 206/µL.

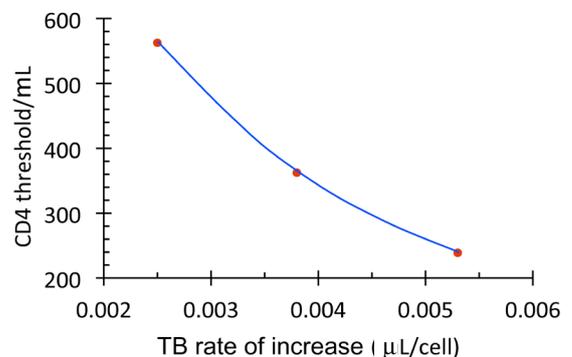

Figure 6. The CD4 threshold, below which ART will lead to an increase in the incidence of TB, as a function of the rate of increase of TB with declining CD4 counts,

## Discussion

We have only considered average values for the initial CD4 counts and for survival even though both vary widely. Since the results are insensitive to the initial CD4 count or to the survival without ART, this is unlikely to affect the conclusions. The only substantial uncertainty arises from the estimate of the rate at which the incidence of TB increases with declining CD4 counts and better data on this would be informative.

In this analysis we assume that the CD4 count declines linearly with time. While this does have some support[16-19] it needs further verification.

The assumption that the extent of the recovery of CD4 counts after the start of ART is independent of the CD4 count at which people start treatment is curious but has strong support from two studies.[4-6] Given the variation in the CD4 counts in HIV-negative people[7] it would be useful to have similar data for other populations.

## Conclusion

Increasing the CD4 count at which people start treatment will always reduce a person's lifetime risk of developing TB. To ensure that this risk is also less than the risk in people who never start ART, and die ten years after they become infected with HIV, HIV-positive people should start ART at a threshold CD4 count above about 350 cells/μL. This threshold is insensitive to their CD4 count before they become in infected with HIV but is sensitive to the rate at which the incidence of TB increases as CD4 counts fall. This analysis also ignores the substantial reduction in HIV prevalence, and therefore of HIV-related TB,[2] that would result from widespread use of ART and it is unlikely that the wide spread use of ART will increase the overall incidence of TB.

## Conclusion

If people start ART very late in the course of their infection this will increase their lifetime risk of TB. Since it will also have little impact on the overall prevalence of untreated HIV it will lead to a long-term increase in the incidence of TB in the population as a whole. If people start ART very early in the course of their infection this will decrease their lifetime risk of TB. If coverage of ART is also high this will have a substantial impact on the overall prevalence of untreated HIV and lead to a long-term decrease in the incidence of TB in the population as a whole.

Our best estimate of the CD4 threshold, which determines whether ART increases or decreases TB incidence in the long term, is about 350 cells/μL. If ART is started at this value of the CD4 count the lifetime risk of TB is about the same as in people who never start ART. However, provide there is high coverage, starting ART at this threshold ART should reduce the overall incidence of HIV by about 50% and the steady state prevalence of HIV by about 30%,[2] which will ensure that the overall incidence of TB falls.